\documentclass{ws-fnl}
\usepackage{cite}
\usepackage[colorlinks=false]{hyperref}

\usepackage{color}
\definecolor{orange}{RGB}{255,152,24}
\definecolor{fuchsia}{RGB}{255,92,255}

\begin{document}

\markboth{F. Turci and  E. Pitard}{Dynamical heterogeneities in a 2-dimensional driven glassy model:
 current fluctuations and finite-size effects (UPoN-2012 Submission}
\title{DYNAMICAL HETEROGENEITIES IN A \\  TWO DIMENSIONAL DRIVEN GLASSY MODEL :
\\ CURRENT FLUCTUATIONS AND FINITE SIZE EFFECTS \\
(UPoN -2012 SUBMISSION)}
\author{FRANCESCO TURCI}
\address{Laboratoire Charles Coulomb \\ Universit\'e Montpellier II and CNRS, 34095 Montpellier, France\\
francesco.turci@um2.fr}
\author{ESTELLE PITARD}
\address{Laboratoire Charles Coulomb \\ Universit\'e Montpellier II and CNRS, 34095 Montpellier, France\\
estelle.pitard@um2.fr}

\maketitle

\begin{history}
\revised{21st July 2012}
\end{history}
\begin{abstract}
In this article, we demonstrate that in a transport model of particles
with kinetic constraints, long-lived spatial structures are responsible
for the blocking dynamics and the decrease of the current at strong driving field. Coexistence 
between mobile and blocked regions can be anticipated by a first-order transition in the large deviation function for the current.
By a study of the system under confinement, we are able to study finite-size effects and extract a typical length between mobile regions.
\end{abstract}


\section{Large deviations in kinetically constrained models}\label{LDF}
\subsection{Large deviations formalism}

The large deviation theory can be viewed both as an extension and  a new framework for developing statistical mechanics \cite{Touchette:2009p4586}, and relies on probabilistic foundations. 
This formalism yields information about the fluctuations
of temporal trajectories in configuration space, in analogy with
the usual canonical thermodynamics approach which gives access to the fluctuations of observables at equilibrium
such as the energy.

This approach can be extended to  out-of-equilibrium systems, 
and in particular can be applied to Markovian dynamics where 
a well defined steady state exists: equivalents
 of free energies and entropies can be defined for observables
  extensive in time and can be computed either analytically (for very simple toy models) or numerically \cite{Lecomte:2007p16723}. The  tails of the distribution and the  fluctuations of a given observable can then be  quantified and reflect its sensitivity to the initial conditions. 

In a very general way, for a given process having a well defined steady state, if $O$ is a time-extensive observable we can write the associated partition function 

\begin{equation}
Z_O(s,t)=\sum_{histories} {\rm Prob(history)} e^{-sO({\rm history})},
\end{equation}
 where a history is the time series of configurations $\left\{C_{1}, C_{2},\dots,C_{t}\right\}$ and $s$ is a transformation variable conjugated to the observable (via a Legendre transform), so that histories having a non typical value of $O$  are selected 
according to the value of $s$. $O$ can typically be the activity (the time integrated number of particle moves) or the integrated current.
 $s$ can be
viewed as a  \textit{chaoticity temperature} since it allows to select histories 
having either more and more rare events (for $s$ positive) or more and more frequent events 
(for $s$ negative).
  
In the long time limit, the partition function behaves like
\begin{equation}
Z_O(s,t)
       \simeq \exp\left\{t\psi_{O} (s)\right\},
\end{equation}
where $\psi_{O}$ is called the \textit{large deviation function} associated with the observable $O$. 

The function $\psi_{O}(s)$ is the cumulant generating function
of $O$, since all the cumulants of the underlying distribution can be obtained by means of derivation. For example the stationary value of $O$ is simply given by 
\begin{equation}
\lim_{t\rightarrow \infty}
\frac{\langle O(s,t)\rangle}{t}=-\psi'_O(s)|_{s=0}
\end{equation}
where the average is taken over the ensemble of histories. Similarly, higher order cumulants are obtained from higher order derivatives.

\subsection{Application to kinetically constrained models}
In the context of the study of the glass transition and the jamming transition, 
the slow, frustrated dynamics of real glasses has been modeled under
 strong simplifications by particles/spin models where the moves/spin flips 
 have some dynamical restrictions. Such toy models (referred to as \textit{kinetically constrained models} or KCM \cite{Ritort:2003p99,Cancrini:2010p10535}) allow to explore elementary properties of glassy systems, such as long relaxation times, aging dynamics, anomalous diffusion etc. with a minimum number of ingredients. 

Given their relative simplicity, these models have 
been considered as good candidates in the field of disordered systems 
for applying the concepts coming from the thermodynamics of histories. 
Many of the glassy properties related to the slowing down of the dynamics 
when  the temperature decreases (or, equivalently, when the density of particles
increases) are thought to be related to the formation of clusters 
of blocked particles slowly relaxing in the bulk, 
keeping the system amorphous on long time scales but collectively organized
 on shorter time scales: these patches of clustered particles,
  moving cooperatively in a non diffusive manner are often called dynamical heterogeneities
  \cite{Kob:1997p8626, Bertin:2005p12691, Heussinger:2010p12749}. 
  A  natural time extensive observable allowing to quantify the strength
   of such blocking and clustering processes is the total activity K(t), defined
    as the number of moves or spin flips from a 
     reference time $t_{0}=0$ to time $t$: mobile patches will increase the total activity while blocked regions will keep it constant.

The numerical study of KCM like the Fredrickson-Andersen model \cite{Fredrickson:1984p10646}, 
the Kob-Andersen 2d lattice gas \cite{Kob:1993p3432} etc. via the thermodynamics of histories 
can be made possible thanks to biased ensemble
 algorithms \cite{Giardina:2006p4759,Tailleur:2008p4778,Giardina:2011p16490} : 
 either in continuous or in discrete time, it is possible to make evolve a 
 certain population $N$ of different clones or replicas of the system 
  and make them perform a biased dynamics so that well defined 
  fluctuations of the activity are explored for a given value of $s$; 
  such biased dynamics allows the direct computation of 
  the cumulant generating function associated to the activity $\psi_{K}(s)$,
   which by construction is  analogous to the Helmoltz free energy of equilibrium canonical ensembles. 

In the case of kinetically constrained models,
 the derivative of $\psi_{K}(s)$  is discontinuous at $s=0$: 
 this first order dynamical phase transition
  has been interpreted \cite{Garrahan:2007p16652} as the signature 
  of separation in the space of possible histories between active histories,
   where particles manage to escape from cages, and inactive ones, 
   where particles are blocked by the dynamical restrictions. 
   In particular, the first order transition at $s=0$ (unbiased dynamics) 
   is interpreted as the coexistence of such histories in  real space dynamics, this vision being compatible with the cage dynamics for which KCM have been proposed as model glass formers.

What happens if the system is driven by an external force? 
Does the heterogeneous behavior persist in terms of 
the large deviation approach? In order to address these questions,
 we consider a simple driven KCM model for granular materials.


\section{The Kob-Andersen model in presence of a drift}
M. Sellitto has recently proposed \cite{Sellitto:2008p3255} an out of equilibrium variation of the original Kob-Andersen model, introducing a basic mechanism for mimicking an external shear over a granular fluid. This model can also be reinterpreted as the kinetically constrained version of a 2d Asymmetric Simple Exclusion Process (ASEP).

sOn a 2d regular square lattice of size $L\times L$,
 the Kob-Andersen model is a kinetically constrained model \cite{Kob:1993p3432}
  where hard-core particles can move to any of the nearest-neighbour empty
   sites under the condition of having no more than 
   two neighbouring particles before and after the move. The motion
    is isotropic, and the system is ergodic at any density,
     even if the relaxation times rapidly increase at high densities.

Under periodic boundary conditions,
 it is possible to observe the set up of a
  macroscopic current as soon as an external bias 
  is introduced. The simplest way to do it is to impose 
  a uniform,
   unidirectional external field: following a 
   discrete time dynamics, particles can hop in a given 
   direction with probability $p=\min \{1,\exp(\vec{E}\cdot d\vec{r})\}$
    where $d\vec{r}$ is the displacement vector from cell to cell.
     Combining these rules of motion and the Kob-Andersen kinetic
      constraint leads to the asymmetric exclusion processes with constrained dynamics \cite{Sellitto:2008p3255}.

\begin{figure}[t!]
\begin{center}
\includegraphics[width=\columnwidth]{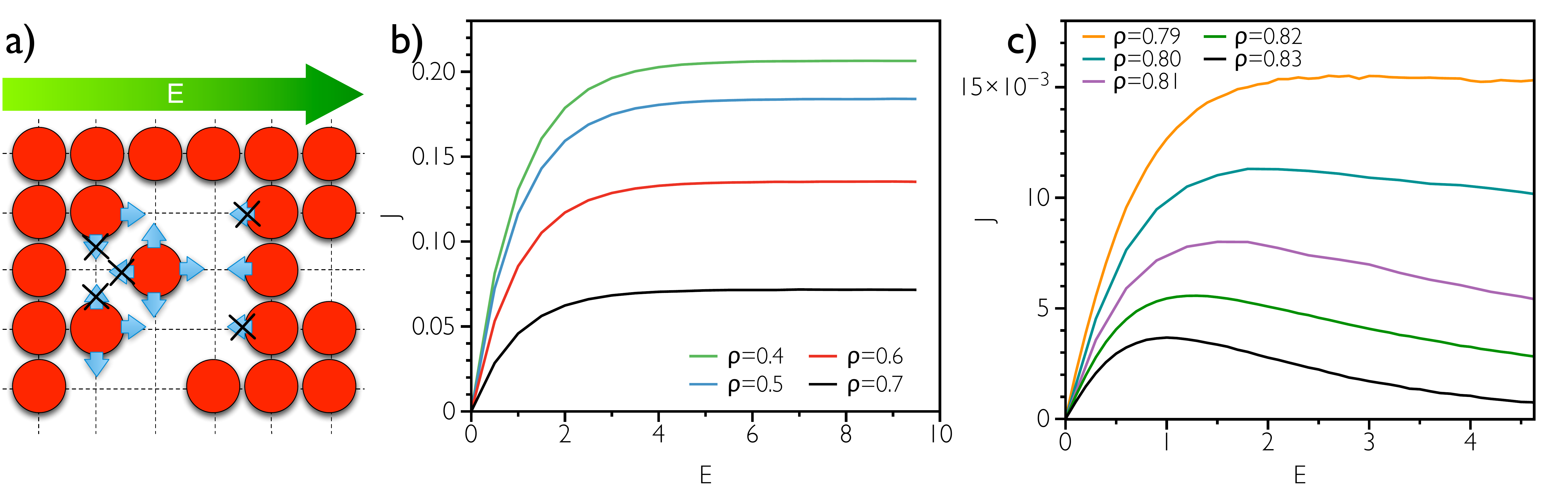}
\caption{(color online) \textbf{a)} Dynamic rules for the Kob-Andersen model
 in presence of a drift ; particles (red circles) 
 can move and occupy empty sites if they have at least 2 empty 
 neighbors before and after the move; an external field $E$ 
 acting on the whole volume is imposed in order to bias the dynamics in a given direction. \textbf{b)}Current vs field intensity relation for a system of linear size  $L=100$ for different values of the density; at low densities $J(E)$ is a monotonic function, while at densities above $\rho_{c}\sim 0.79$ \textbf{c)} the relation becomes non monotonic.}
\label{model}
\end{center}
\end{figure}

At low densities the overall dynamics is well modeled by an effective 2d Asymmetric Simple Exclusion Process, without the Kob-Andersen constraints:
 the current is then proportional to the difference $1-e^{-E}$,
  which measures the
   difference between the hopping rates in the positive and negative field direction.
    This leads to a monotonic behavior for the current vs field relation $J(E)$,
     which quickly saturates as the field intensity increases.
      We see that  the introduction of the Kob-Andersen
	 kinetic constraints does not lead to any important change in the
	  dynamical processes with respect to normal ASEP (see fig. \ref{model}.b):
	  this  means that the correlations between particles
       movements do not go far beyond on-site hard core repulsion,
        at least for small densities.

However, when the density of particles 
increases, the monotonicity breaks down. This happens approximately 
at density $\rho_{c}\approx 0.79 $ (fig. \ref{model}.c) and the current eventually 
saturates at a finite value at sufficiently large field intensity.
 The presence of a maximal value for the current allows to distinguish 
 two characteristic dynamical regimes: for field intensities 
 smaller than the optimal value $E^{*}(\rho)$ the relaxation times 
 are comparable to what happens in the $E=0$ case, while for $E>E^{*}$ 
 the dynamics slows down exponentially with the external field intensity \cite{Sellitto:2008p3255}.

Here we focus on the dynamics in the high density, strongly correlated regime, with particular attention to the sensitivity to initial conditions, the bimodality of the distributions of certain dynamical observables (in particular the activity and the current) and their scaling with the system size.

\subsection{Large deviations for the activity and the current}

Using the large deviation formalism introduced 
in section \ref{LDF} both the activity 
(the number of microscopic movements in a whole history) 
and the total integrated current have been studied \cite{Turci:2011p15998}. 
From this analysis, it results 1) that a first-order dynamical transition 
exists for the activity for any value of the external field, 
being the signature of the presence of the kinetic constraints and 
2)  the current shows a first-order
transition between an active and an inactive phase only in the limit of very strong fields.
 The interpretation of this phase transition occurring 
 in the strong field regime (where the current decreases when the field increases) 
 can be given by the accurate analysis of the changes that occur in configuration space 
 when the external field is increased.

The first order transition in the current large deviation function has to be interpreted as the coexistence of flowing and non flowing histories for the system, and a strong sensitivity to initial conditions. The flowing and non flowing histories phase coexistence is the signature of a  bimodal distribution of the current in the thermodynamical limit.
In practise (at $s=0$), there is only an exponentially small number of trajectories that make the system blocked or with very small current; these rare histories are better sampled in  the biased dynamics (at $s>0$). We shall see in the following that this picture can be improved by a  finite size analysis in the configurations space.

\begin{figure}[h!]
\begin{center}
\includegraphics[width=0.6\textwidth]{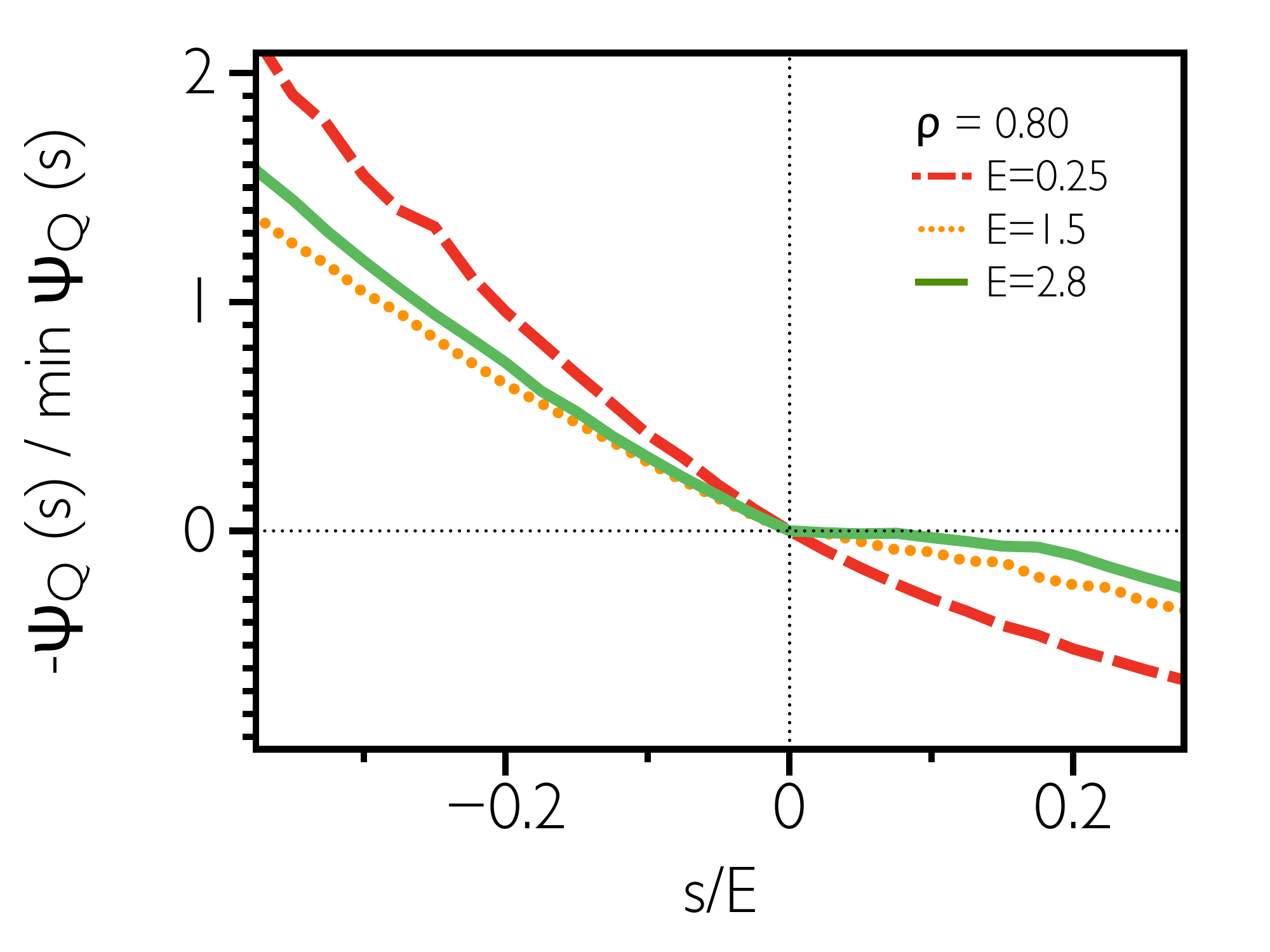}
\caption{
Numerical evaluation of the large deviation function for the current $\psi_{Q}(s)$ for a system of linear size $L=60$ at density $\rho=0.8$ \cite{Turci:2011p15998} in the vicinity of $s=0$. The discontinuity in the derivative of the large deviation function at the origin is clearly visible for strong fields.}
\label{ldfANDpdf}
\end{center}
\end{figure}
Indeed the origin of active and inactive patches resides on the microscopic dynamics of the rearrangements of particles and the limitations due to the coupling of the kinetic constraint with the increase of the external field: these two ingredients are responsible for a complete blockage of the dynamics in the limit of infinite lattice sizes and fields. Such microscopic rearrangements give rise to the transversal structures composed by filled and empty regions (that we have called {\em{domain walls}}) which allow for a corse-grained interpretation of the dynamics in terms of a patchwork of active and inactive regions of quantifiable size.

\subsection{Blockage dynamics}
The variation of the external field intensity triggers modifications in the real space structure of the system.

\begin{figure}[h!]
\begin{center}
\includegraphics[width=\textwidth]{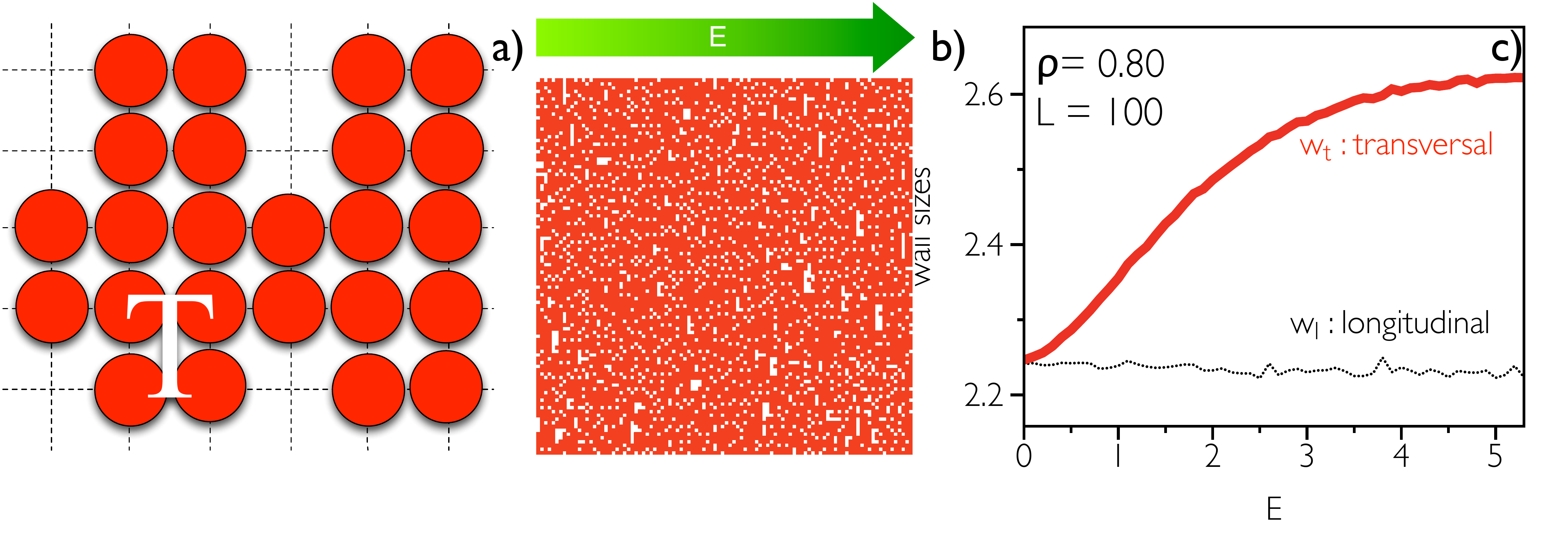}
\caption{a)The square lattice Kob-Andersen models 
admits some frozen configurations composed 
by T-juctions of double-width occupied bars (particles in red). 
While these configurations are rare in the undriven case ($E=0$), 
the application of an external field ($E=5$ and $\rho =0.8$ in  panel b) )
 enhances the probability of such configurations,
  where the residual mobility is assured by localized defects (holes) 
  in the vicinity of the junctions.
   c) For a model of linear size $L=100$ at density $\rho = 0.8$,
    connected clusters of holes
     allow the definition of a transversal and longitudinal length,$w_t$ and $w_l$ respectively.
      Transversal walls of holes (thick red line)
       grow when the external field increases
        while in the longitudinal direction 
	(dotted line) weak field effects are measured.
	 Average values have been computed over 100 different histories.}
\label{Tcrossing}
\end{center}
\end{figure}

Once the external field is turned on,
 it is possible to observe \cite{Turci:2011p15998}
  the formation of structures in the transverse direction: 
  clusters of holes surrounded by bars of particles
   that eventually form T-junctions at the edges.
    These areas are not permanently blocked:
     defects (neighboring holes) are present
      so that the structures may relax,
       but the time needed for the relaxation
        is largely increased with respect to the undrifted case (fig. \ref{Tcrossing}).
	 It is possible to measure the extension
	  of the connected clusters of holes in
	   the longitudinal and transversal direction.
	    It appears (as discussed in \cite{Turci:2011p15998})
	     that the transversal length is the most relevant
	      and defines a \textit{walls' length}.
	       These \textit{walls of holes} block the dynamics: 
	       their growth as a function of the external field
	        is directly linked to the non monotonic behavior of the current as a function of the field.
\section{Finite size analysis}
\subsection{Bimodality of the current distribution in real space}
The role of transversal walls in the crossover from a monotonic to
 a non monotonic $J(E)$ relation allows to shed light on the sensitivity to initial conditions and the bimodal nature of the distributions of current and activity.

We have chosen to modify the protocol of our
 simulations in order to explore the role of relevant lengthscales, 
 in particular in the transversal direction. We still keep 
 the torus geometry for our system (periodic boundary conditions in both directions),
  but we now differentiate the transversal dimension $H$ from the longitudinal dimension $L$:
   in order to keep the same statistics for particle moves for different system sizes, we rescale $H$ in order to keep the product $H\times L\sim 60^{2}$ constant, hence the number of particles is kept constant. 

Reducing the transversal size $H$ makes 
the transversal blockages extension critical for the current flow. 
If the blocked regions are of size comparable to the transversal dimension $H$, 
these regions slow down the flow in the system, leading it to an absorbing frozen 
state where no net flow is produced. The realization of such configurations 
depends both on the initial conditions and on the history. 

\begin{figure}[h!tbp]
\begin{center}
\includegraphics[width=0.63\textwidth]{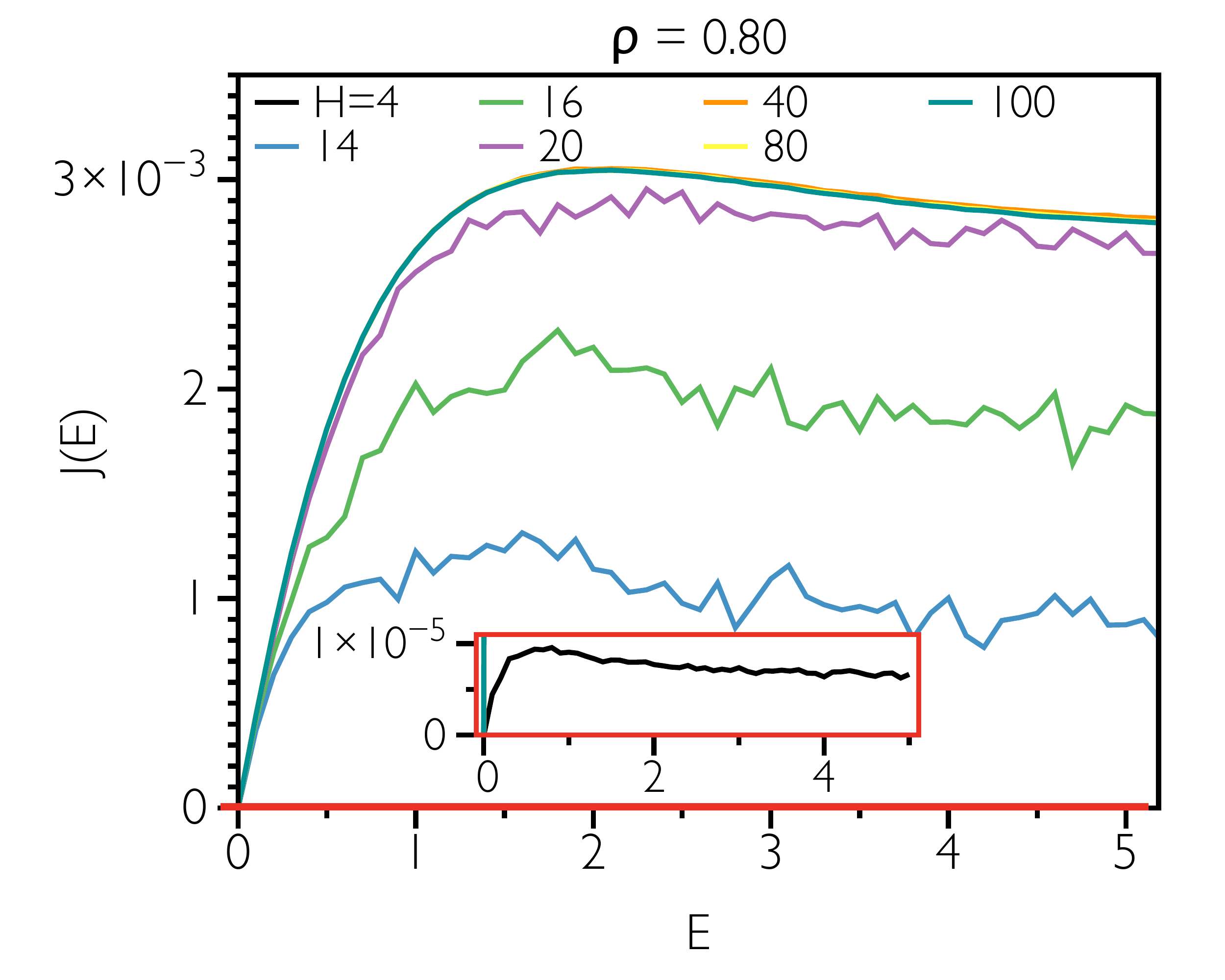}
\caption{Current vs field relation for different values of the thickness of the system $H$ 
at density $\rho =0.80$. Decreasing the transversal dimension 
($H\lesssim 30$) increases the current fluctuations; although 
all averages are performed over 150 different histories,
the noise of the data clearly increases when $H$ decreases.}
\label{reducingH}
\end{center}
\end{figure}

\begin{figure}[t!]
\begin{center}
\includegraphics[width=0.65\textwidth]{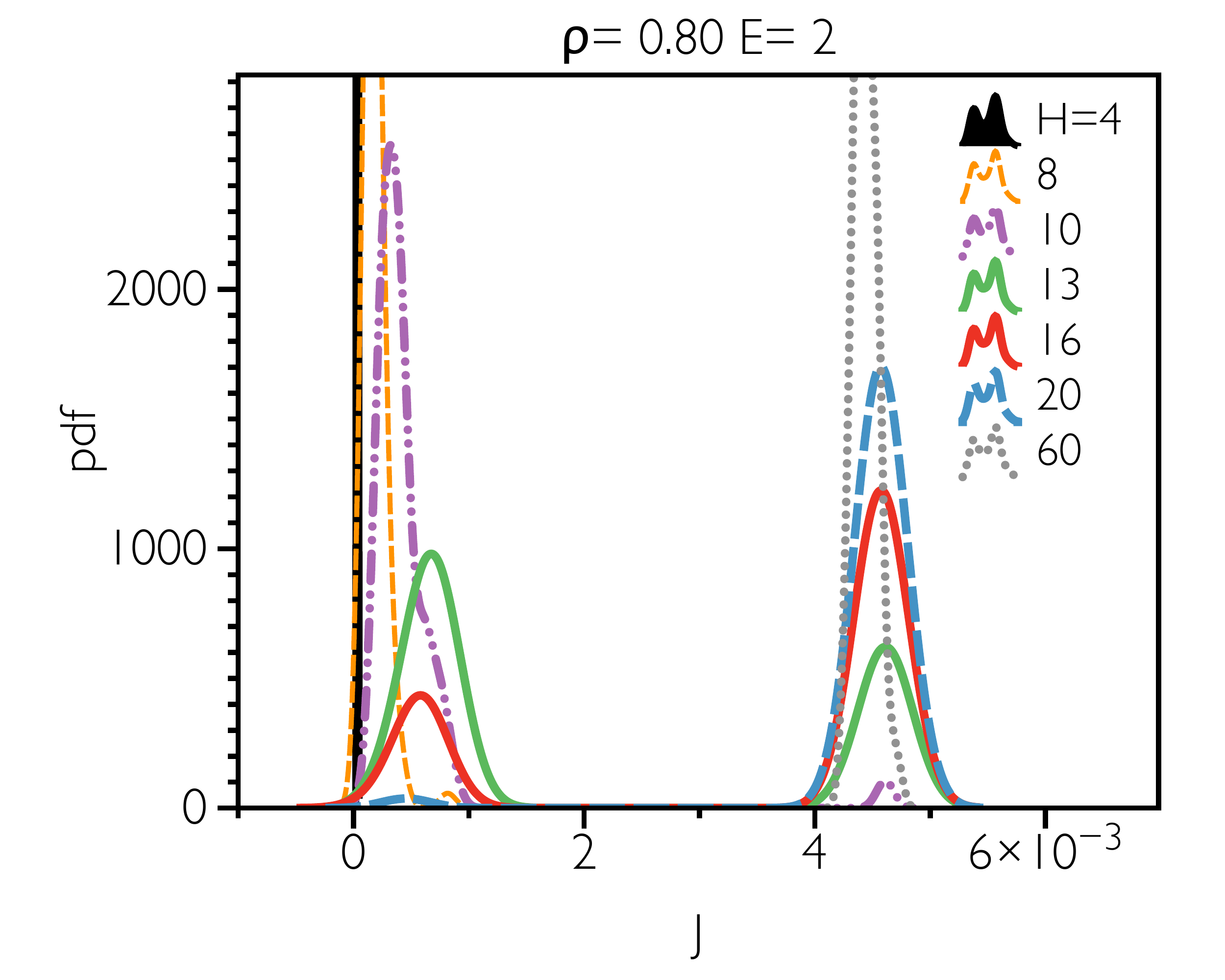}
\caption{Current distributions for strong fields ($E=2$) at density $\rho = 0.80$ 
for various transversal dimensions $H$. 
Large systems have unimodal distributions peaked around a positive value of the current, 
while very small ones are peaked around zero. 
Intermediate sizes show bimodal distributions with different amplitudes of the peaks. 
The distributions are computed as kernel density functions over 100 different histories.}
\label{currdistrE2}
\end{center}
\end{figure}

Moreover, the growth of the fluctuations 
of the current is the signature of an important change in the shape of
the distribution of the current. As shown in figure \ref{currdistrE2}, large systems correspond to unimodal current distributions peaked around a well defined positive value. When the transversal size is reduced (to $H=20$) a second peak appears in the vicinity of $J=0$, whose relative amplitude increases when the system size is reduced. For very small $H$, this second peak at zero current becomes dominant, meaning that the histories corresponding to some non-zero flow become exponentially unlikely.
 
\subsection{Dynamical correlation length}
 In order to better highlight the relation between structural properties 
 and the dynamics, one can try to couple the changes in the dynamical properties as a function of the external field to the  variation of the transversal system size $H$.
To do so, we can use the definition of relaxation times 
given in \cite{Sellitto:2008p3255} as the integral 
\begin{equation}
\tau_{rel}(E,\rho,H)=\int_{0}^{\infty} \phi (t) dt,
\end{equation}
where $\phi(t)$ is the persistence function
 quantifying  the probability that the occupation 
 variable of a lattice site has never changed between time 0 and t. 
As noted in \cite{Toninelli:2004p12011,Toninelli:2005p16725}
 for the undrifted 2d Kob-Andersen model (which is ergodic for all densities below $1$) it is possible to define a characteristic length scale 
\begin{equation}
\Xi_{\rho}\sim \exp\left\{		\frac{\pi}{18(1-\rho)}\right\},
\end{equation} 
which gives an 
 estimate of the spacing between mobile elements: 
 in this sense it can be viewed as a dynamical correlation length. 
 A system much larger than this length scale will have a well-defined
relaxation dynamics. Below this length, 
the system may break into blocked patches that do not relax.
In the case of a macroscopic flow, 
this breakdown of the system is clearly a limiting factor:  
no macroscopic current can pass through, reducing the current contribution to zero. 
Moreover, the smaller the system is, the longer will be the relaxation time, 
because more and more areas will be frozen in their initial state. 
We compare the relaxation times of the system to its linear dimensions in order to extract the crossover length $\Xi_{\rho}$. In our case, this length will be dependent on the external field intensity $E$.
\begin{figure}[h!tbp]
\begin{center}
\includegraphics[width=0.7\textwidth]{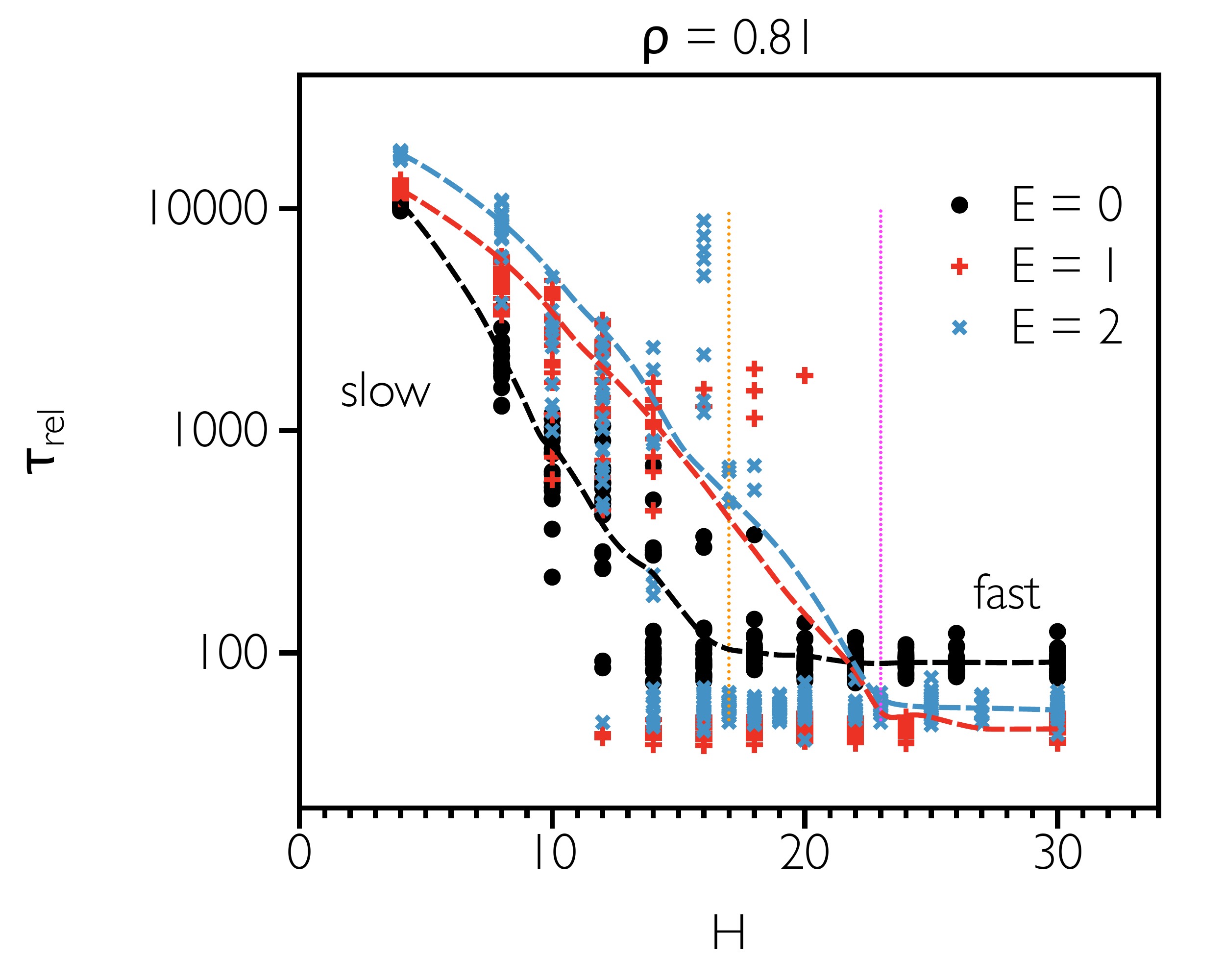}
\caption{(colors online). Scatter plot of the relaxation times computed 
for different system sizes at different values of the external field.
 Each point corresponds to a unique history, so that for every couple  $(H,E)$
  20 different histories are considered. 
  One can easily distinguish two groups of points: 
  slow trajectories and fast ones, here distinguished by a gray dashed line.
   LOESS (LOcally Estimated Scatterplot Smoothing) best fits are also shown as dashed lines
    in order to illustrate the behavior of the average value of the relaxation times as a function of $H$. 
In the case of $E=0$ (black curve) the kink in the average 
corresponds to the value of the predicted critical length $\Xi_{0.81}\sim 17$
 ({\color{orange} orange} dotted line), which rapidly grows for
  stronger fields and seems to saturate at a value
$\Xi_{0.81}^E\sim 23$({\color{fuchsia} fuchsia} line).}
\label{tau_rels}
\end{center}
\end{figure}

The numerical study shown in figure \ref{tau_rels} 
demonstrates how the sensitivity to initial conditions 
and the existence of a critical length 
are combined and depend on the intensity of the external field. 
In fact, while for sufficiently large systems one typical, 
almost constant relaxation time is well defined for any value of the external field,
reducing the transversal dimension $H$ 
leads to a rapid increase of the relaxation times at a certain $H^{*}(\rho, E)$;
 eventually  very small systems are practically arrested. 
 Moreover, around the critical length $H^{*}(\rho, E)$ 
 the distribution of the relaxation time splits in two parts, 
 becoming bimodal:  while a fraction of histories 
 is characterized by the same relaxation time as large thermodynamic systems,
  a  fraction of histories
  jumps to much higher relaxation times, getting trapped and slowed down by the formation of blocked patterns.

If one takes the average relaxation times, 
one can observe that the kink corresponding 
to the change of slope of the $\log \langle\tau_{rel}(H)\rangle$ 
in the case of $E=0$ approximately corresponds to the critical length 
scale $\Xi_{\rho}$ predicted in \cite{Toninelli:2005p16725}, so that we have
 $H^{*}(\rho,0)\sim\Xi_{\rho}$. 
 The numerical simulations then suggest that when the field increases this critical length correspondingly increases: this is coherent with the picture provided before concerning the formation of new correlated structures, slowing down the dynamics and making harder and harder for particles to escape from their initial configuration.
\section{Conclusions}
We have discussed the dynamical implications 
due to an externally driven jamming transition for a simple 2d model. 
Dynamical heterogeneities have been related to the 
realization of rare rearrangements driven by the external field and 
the intrinsic bimodal nature of the distribution of current. 
A finite size effects analysis allowed to explore in real space such bimodality, 
previously signaled by a study in terms of large deviations functions.
\bibliographystyle{ws-fnl}
\bibliography{references}
\end{document}